\documentclass[11pt]{article}

\usepackage[preprint]{acl}

\usepackage{times}
\usepackage{latexsym}

\usepackage[T1]{fontenc}

\usepackage[utf8]{inputenc}

\usepackage{microtype}

\usepackage{inconsolata}

\usepackage{graphicx}

\usepackage[inline]{enumitem}
\usepackage{amssymb}
\usepackage{amsmath}
\usepackage{booktabs}
\usepackage{makecell}
\usepackage{todonotes}
\usepackage{multirow}
\usepackage{listings}

\newcommand{\tabjoinbench}{\textsc{TabJoinBench}}

\title{\tabjoinbench: A Benchmark for Joinable Table Discovery}

\author{
\textbf{Sandipan De}
\quad
\textbf{Jin Wang}
\quad
\textbf{Vivek Gupta}
\\[2pt]
Arizona State University
\\[4pt]
\texttt{\{sandipan, jinwang18, vgupt140\}@asu.edu}
}

\begin{document}
\maketitle
\begin{abstract}
Join discovery aims to identify tables from large data repositories that can augment a query table with complementary information, enabling downstream tasks such as data exploration, feature engineering, and business intelligence. Although numerous join discovery methods have been proposed, existing studies rely on method-specific benchmark construction, making reproducible and fair comparison difficult. We present \textbf{\tabjoinbench}, a benchmark for evaluating join discovery methods across semantic, relational, and hybrid data lake scenarios. {\tabjoinbench} constructs query--candidate pairs using source-specific validation strategies, systematically introduces structural, representation, and semantic changes through composable perturbations while preserving reliable ground truth. We evaluate representative join discovery methods spanning set-based, feature-based, and learned approaches, together with general-purpose language-model embedding baselines, and publicly release the processed datasets, ground-truth annotations, and generation pipeline to facilitate reproducible evaluation and future research.
\end{abstract}

\section{Introduction}
Data lakes enable organizations to store large collection of data originating from diverse sources, including enterprise repositories containing internal organizational data, data from partner organizations supporting inter-organizational collaboration, and open data or web repositories providing reference and contextual data. While these repositories provide significant opportunities for analytics, their heterogeneous nature makes discovering relationships between datasets a fundamental challenge. Common relationships between datasets include duplication (equivalent datasets), containment (one dataset is a subset of another), unionable (datasets contain disjoint but schema-compatible records that can be combined to improve coverage), and complementarity (datasets contain different but related attributes that can be integrated through joins).

In this work, we focus on the join discovery task that aims to identify datasets which can be meaningfully joined given one or more shared attributes and a target dataset. Unlike traditional database systems where key relationships are explicitly defined, data lakes typically lack schema constraints, consistent naming conventions, or standardized value representations. As a result, two datasets describing the same real world entities may use different schemas, different value representations, or different semantic concepts to represent equivalent information. These challenges have motivated extensive research on scalable join discovery methods based on set similarity \cite{zhu_2019_josie, dong_2021_pexeso}, learned representations \cite{dong_2023_deepjoin, koutras_2025_omnimatch, khatiwada_2025_tabsketchfm} or featured-based retrieval techniques \cite{bogatu_2020_d3l, maynou_2026_freyja}.

Despite the advancement of join discovery algorithms, their experimental evaluation remains fragmented. These methods commonly evaluate on public data repositories such as WDC Web Table Corpora \cite{lehmberg_2016_wettables}, WikiTables \cite{bhagavatula_2015_wikitables}, or Open Government datasets, but rely on method specific preprocessing. Although the original datasets are publicly available, the processed datasets, benchmark queries and ground-truth annotations are often not released, making exact reproduction and fair comparison difficult. More recent benchmark efforts \cite{srinivas_2023_lakebench, deng_2024_lakebench} substantially improve reproducibility by releasing datasets, queries and ground truths. However, these benchmarks are primarily constructed by vertically and horizontally partitioning datasets while preserving one or more shared columns, thus evaluating equi-joins between closely related tables. Although they capture the scale of data lakes, they fail to reflect the semantic diversity, data quality issues, and ambiguous join relationships observed in practice. Moreover, they generally assume that the shared column is a valid join key without verifying its semantic correctness.

To address these gaps we present \tabjoinbench, a benchmark that represents three distinct join discovery scenarios. The benchmark is constructed from: 
\begin{enumerate*}[label=(\alph*)]
    \item WikiTables \cite{bhagavatula_2015_wikitables}, providing semantically related table pairs identified through Wikidata \footnote{\url{https://www.wikidata.org/}} entity overlap
    \item MMQA \cite{wu2025mmqa}, providing relational equi-joins extracted from ground truth SQL annotations for Text-to-SQL tasks
    \item NYC Open Data \footnote{\url{https://opendata.cityofnewyork.us/}}
\end{enumerate*}, representing realistic data lake scenarios through key-aware partitioning.
Together the benchmark captures complementary characteristics of web tables, relational databases and data lakes.

The main contributions of this work are:
\begin{itemize}
    \item We propose a benchmark construction methodology that establishes validated seed pairs and systematically transforms them into realistic data lake through composable perturbations while preserving reliable ground truth.
    \item We release {\tabjoinbench} along with its datasets, queries, ground truth annotations, and the generation pipeline to facilitate reproducible evaluation of join discovery methods.
    \item We perform a comprehensive evaluation of representative join discovery methods. Our study compares set-based, feature-based, learned, and embedding based approaches and analyze their robustness under different types of join semantics and perturbations.  
\end{itemize}

\section{Preliminary: Join Discovery}
Let the data lake $\mathcal{L}$ be a collection of tables $\{T_1, T_2, \ldots, T_n\}$, where each table $T_i = \{C_i^1, C_i^2, \ldots, C_i^{m_i}\}$ consists of $m_i$ columns. Similarly, let the query set $\mathcal{Q}$ be a collection of query tables $\{T_q^1, T_q^2, \ldots, T_q^p\}$. 

Given a query table $T_q\in\mathcal{Q}$ and one of its columns $C_q^k$, the objective of join discovery is to identify columns from the data lake that can be meaningfully joined with the query column. A join discovery method defines a relevance function
$R(C_q^k,C_i^j)\rightarrow\mathbb{R}$
that estimates the relevance of a data lake column $C_i^j$ with respect to the query column $C_q^k$. Let $\mathcal{G}$ denote the set of ground-truth joinable columns for the query column $C_q^k$. The method calculates the relevance such that, for every column $C_x\in\mathcal{G}$ and every column $C_y\notin\mathcal{G}$,
$R(C_q^k,C_x) > R(C_q^k,C_y).$



The relevance function $R(\cdot)$ is method dependent and may be computed using one or more characteristics of the query and candidate columns, including value overlap or similarity, schema information, statistical properties, learned representations, or contextual information from neighboring columns or the entire table. Its interpretation also depends on the type of join relationship: equi-joins require exact or near-exact value correspondence, fuzzy joins relax this to allow syntactic variation, and semantic joins further generalize to columns representing the same underlying concept despite limited lexical overlap. $R(\cdot)$ thus serves as a general abstraction over these notions of joinability.

\begin{figure*}[!ht]
\centering
\includegraphics[width=.98\textwidth]{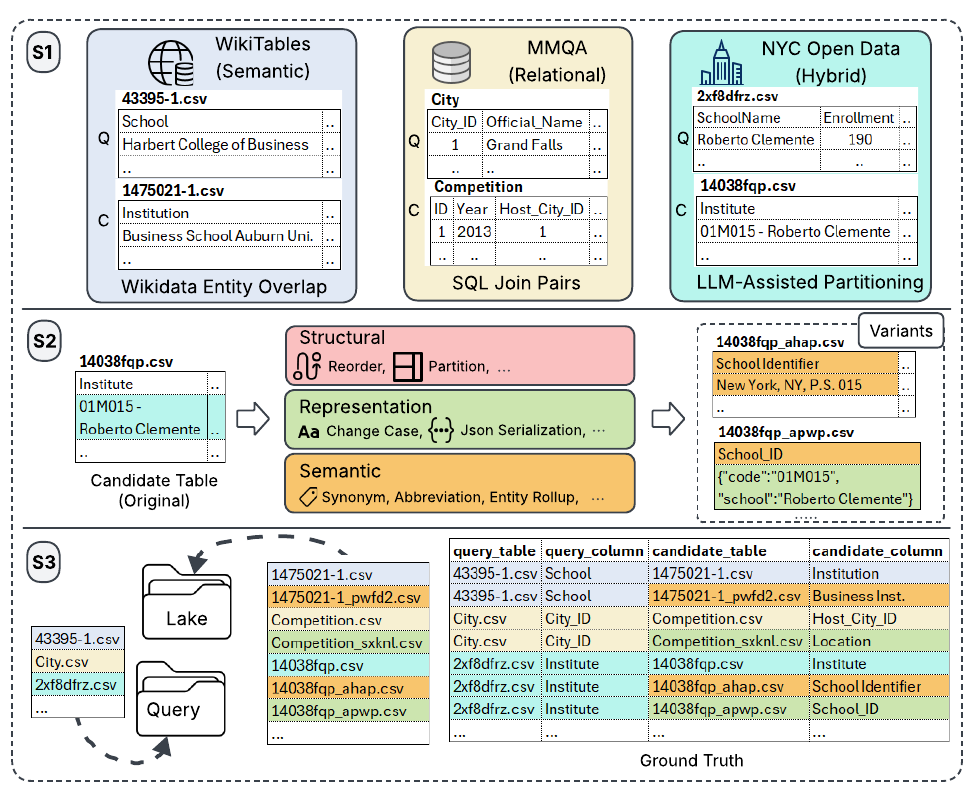}
\caption{Overview of \tabjoinbench{}. Seed join pairs are generated from three complementary data sources (S1), candidate tables are transformed using composable perturbations while preserving the underlying join relationships (S2), and the resulting query set, heterogeneous data lake, and ground-truth annotations are assembled into the final benchmark (S3).}
\label{fig:construction_pipeline}
\end{figure*}

\section{\tabjoinbench}
\subsection{Overview}
{\tabjoinbench} is a benchmark for evaluating join discovery methods across diverse data characteristics and join semantics. Its construction is guided by three design principles. \textbf{First}, it should support multiple notions of joinability, including equi-joins, fuzzy or approximate joins, and semantic joins. \textbf{Second}, the query-candidate pairs should derive from meaningful join relationships, rather than arbitrary shared attributes. \textbf{Third}, the generated lake should capture heterogeneity through controlled and composable perturbation reflecting transformation pipelines in a realistic data lake.

Figure~\ref{fig:construction_pipeline} illustrates how the proposed design principles are realized. Stage~1 supports the first two design goals by generating validated seed pairs from complementary data sources. Stage~2 supports the third design goal by introducing heterogeneity through composable perturbations. Stage~3 constructs the final benchmark.


\subsection{Data Sources and Benchmark Families}
\label{subsec:data_sources}
{\tabjoinbench} comprises three complementary benchmark families, each representing a different join discovery scenario. Instead of relying on a single data source, it combines web tables, relational databases, and open data repositories to capture diverse data characteristics.

\paragraph{WikiTables.}
WikiTables \cite{bhagavatula_2015_wikitables} is a corpus of approximately 1.6 million tables extracted from the English Wikipedia. The corpus contains approximately 30 million hyperlinks referencing corresponding Wikipedia pages. These hyperlinks provide entity-level annotations that can be mapped to knowledge-base entities (e.g., Wikidata QIDs), enabling semantic relationships between tables beyond lexical or value overlap. This makes WikiTables suitable for constructing a benchmark for \emph{semantic join} discovery.

\paragraph{MMQA.}
MMQA \cite{wu2025mmqa} is a multi-table question answering benchmark constructed over Spider \cite{yu-etal-2018-spider} database. Each benchmark instance consists of a natural language question, the set of tables required to answer the question, and the corresponding ground-truth SQL query. As the ground-truth SQL query explicitly specifies the join conditions, the equi-join relationships can be derived directly from the query. These characteristics make MMQA suitable for constructing a benchmark for \emph{equi-join} discovery.

\paragraph{NYC Open Data.}
NYC Open Data is a government data portal containing approximately 3000 datasets published by different New York City agencies. Unlike WikiTables, where join relationships are primarily semantic, and MMQA, where joins are defined through relational queries, NYC Open Data contains both identifier-based and textual join attributes. Thus, it allows benchmark construction with \emph{both equijoins and semantic joins}. In addition, the repository consists of a relatively small number of substantially larger datasets than typical web tables, making it well suited for evaluating join discovery methods under realistic data lake settings.

\subsection{Seed Generation and Validation}
\label{subsec:seed_generation}
The three benchmark families exhibit distinct characteristics and therefore require different strategies for constructing seed query--candidate pairs. 

\paragraph{WikiTables Subset.}
The WikiTables subset is constructed from the original WikiTables corpus by identifying semantically related table pairs through shared Wikidata entities. Tables containing fewer than 10 data rows or fewer than 5 columns are excluded, as they provide limited evidence for join relationships. Tables containing more than 75 columns and tables with multi-row headers are also excluded to bound the benchmark to tables with manageable structural complexity. Hyperlinks embedded within the retained tables are resolved to their corresponding Wikidata entities (QIDs), providing entity-level annotations for the table cells.

Join relationships are identified through shared entities occurring within table columns. To reduce spurious matches, entities appearing in fewer than two tables are discarded, while entities appearing in more than fifty tables are ignored as they typically correspond to generic concepts with limited discriminative value. Furthermore, candidate column pairs are required to share at least five common entities before being considered potential joins. These constraints substantially reduce the search space while preserving sufficient semantic relationships.

The resulting query--candidate pairs are further refined to retain only textual join attributes. Columns having a constant character length are excluded because they typically represent fixed-width codes stored as text. Columns with an average character length below five characters are removed as they generally correspond to short categorical values, whereas columns with most common value length exceeding 75 characters are discarded since they primarily contain free-text descriptions. Finally, both the query and candidate column names are required to contain at least two characters to eliminate unnamed or malformed columns.

The remaining query--candidate pairs were prioritized for manual verification using a heuristic score that combines four complementary signals: entity overlap, semantic similarity between table contexts, semantic similarity between the columns, and schema complementarity\footnote{The score is computed as a weighted sum of entity overlap, context similarity, value similarity, and schema complementarity with weights 0.2, 0.2, 0.2, and 0.4, respectively.}. The first three signals reward semantically related table pairs, while the last penalizes highly similar schemas, as such pairs are often duplicate or unionable tables that provide limited additional information when joined.

Applying the heuristic yielded 721 high-confidence pairs with a score of at least 0.9. To evaluate the the scoring function beyond the high-confidence region, an additional 262 pairs were selected for inspection. In total, 983 candidate pairs were manually classified as \emph{joinable}, \emph{unionable}, \emph{subset/duplicate}, or \emph{undetermined} (protocol in Appendix~\ref{appendix:protocol}). Among these, 543 pairs were identified as meaningful joins and retained as the seed query--candidate pairs.

\paragraph{MMQA Subset.}
The MMQA subset is derived from the original MMQA benchmark through parsing the ground-truth SQL. Specifically, the join predicates appearing in the \texttt{JOIN} clauses are parsed to identify the participating tables and their corresponding join columns. When multiple join predicates are present within a query, each participating table pair is treated as an individual seed query--candidate pair. Duplicate pairs originating from multiple questions are removed, resulting in 574 unique seed query--candidate pairs.

\paragraph{NYC Open Data Subset}
Unlike MMQA, the NYC Open Data repository does not provide explicit join relationships between datasets. For this dataset, seed query--candidate pairs are constructed by partitioning individual datasets into two related tables sharing a common join attribute. We filtered the original datasets to include only those containing 10–50 columns and 15–5,000 rows. The lower bounds ensure that the resulting tables contain sufficient schema and data to form benchmark instances. The upper bounds limit the computational cost for benchmark construction. Although real-world data lake contain substantially larger tables, sampling is commonly used in join discovery methods, making this scale suitable for evaluation. After filtering, 620 datasets remain.

For each selected dataset, an LLM is prompted with the dataset title, description, and column names to generate a normalized decomposition of two semantically coherent column groups while preserving exactly one shared column. To guide this decomposition, the prompt instructs the model to identify likely functional dependencies that indicate potential real-world entities. The prompt is provided in Appendix~\ref{appendix:prompt}

Inspired by attribute-affinity-based vertical partitioning algorithm \cite{navathe_1984_vp}, we assess quality of each decomposition by measuring the cohesion between attributes within the generated groups. Traditional vertical partitioning measure attribute affinity through transaction co-access statistics. Since such information is unavailable, we estimate attribute affinity using contextualized column embeddings constructed from the dataset title, description, and column name.

For a column $c$ assigned to group $G(c)$, we calculate an assignment score $\phi$ as ,

\[
\begin{aligned}
\phi(c)=&
\frac{1}{|G(c)|-1}
\sum_{x\in G(c),\,x\neq c}
\mathrm{sim}(c,x) \\
&-
\frac{1}{|\bar{G}(c)|}
\sum_{y\in\bar{G}(c)}
\mathrm{sim}(c,y).
\end{aligned}
\]
where $\bar{G}(c)$ denotes the other group and $\mathrm{sim}(\cdot,\cdot)$ is the similarity between the corresponding column embeddings. A positive score indicates that the column is more semantically related to the columns within its assigned group than those in the other group. The partition score is defined as the average assignment score across all non-shared columns: 
\[
S=\frac{1}{|\mathcal{N}|}
\sum_{c\in \mathcal{N}}\phi(c),
\]
where $\mathcal{N}$ denotes the set of non-shared columns.

Since no ground-truth decomposition exists, the quality of each LLM-generated partition is evaluated statistically. For every dataset, 1,000 random partitions are generated while preserving the same group sizes and the same shared-column. The corresponding partition scores form a null distribution from which the z-score of the LLM-generated partition is computed. Only partitions with a z-score greater than 1.96 are selected for verification. Of these, 82 partitions were randomly selected and manually examined to verify that the generated groups represented meaningful entities and that the shared column constituted a plausible join attribute (guideline in Appendix~\ref{appendix:protocol}). Following manual verification, 70 partitions were retained as seed query--candidate pairs for the NYC Open Data subset.

\subsection{Perturbation Framework}
\label{subsec:perturb_framework}
To evaluate join discovery under a realistic data lake conditions, {\tabjoinbench} augments each validated seed query--candidate pair using a perturbation framework. Each perturbation is implemented as an independent operator, allowing multiple perturbations to be composed into sequential pipelines that simulate the transformations commonly introduced during data cleaning, integration, and exchange. The operators are organized into \emph{structural}, \emph{representation}, and \emph{semantic} categories.

\textbf{Structural perturbations} modify the organization of tables while preserving their information content. These perturbations simulate structural changes introduced in extract-transform-load pipelines, including row and column reordering, vertical and horizontal partitioning, duplicate record insertion, and removal of non-join attributes.

\textbf{Representation perturbations} modify the representation of existing values without changing the underlying meaning. These perturbations emulate formatting and serialization changes commonly introduced during data exchange or import--exports. Example include JSON serialization, contextual prefix, changes in capitalization, missing value insertion, and re-sequencing of identifier column while preserving their one-to-one correspondence.

\textbf{Semantic perturbations} modify the semantic cues available to join discovery methods. These include anonymization and synonym substitution of column names, entity-aware value transformations that produce alternate representations of organizations and products (e.g. abbreviation) , locations and addresses (eg. rollup), and schema changes that merge or split semantically related attributes.

\begin{table}[!ht]
\centering
\small
\begin{tabular}{lll}
\toprule
\textbf{Subset} &
\textbf{Query} &
\textbf{Lake}\\
\midrule
\makecell[l]{WikiTables\\(Semantic)} &
\makecell[l]{Tables: 293\\Avg. Rows: 60.1\\Avg. Cols: 8.1} &
\makecell[l]{Tables: 11932\\Avg. Rows: 55.8\\Avg. Cols: 8.8} \\
\midrule
\makecell[l]{MMQA\\(EquiJoin)} &
\makecell[l]{Tables: 333\\Avg. Rows: 632.0\\Avg. Cols: 5.3} &
\makecell[l]{Tables: 13106\\Avg. Rows: 1270.2\\Avg. Cols: 5.1} \\
\midrule
\makecell[l]{NYC\\Open Data\\(Hybrid)} &
\makecell[l]{Tables: 70\\Avg. Rows: 1015.2\\Avg. Cols: 14.8} &
\makecell[l]{Tables: 2638\\Avg. Rows: 856.8\\Avg. Cols: 10.8} \\
\bottomrule
\end{tabular}
\caption{Summary statistics of the {\tabjoinbench} benchmark, including the numbers of query and lake tables and their average rows and columns.}
\label{tab:benchmark_stats}
\end{table}

\begin{table}[!ht]
\centering
\small
\begin{tabular}{lrrr}
\toprule
\makecell[l]{\textbf{Subset}} &
\makecell[c]{\textbf{No. of}\\\textbf{Queries}} &
\makecell[c]{\textbf{Relevant}\\\textbf{Pairs}} &
\makecell[c]{\textbf{Relevant}\\\textbf{/ Query}} \\
\midrule
\makecell[l]{WikiTables} & 295 & 11,916 & 40.4 \\
\midrule
\makecell[l]{MMQA} & 408 & 13,190 & 32.3 \\
\midrule
\makecell[l]{NYC\\Open Data} & 70  & 2,638  & 37.7 \\
\bottomrule
\end{tabular}
\caption{Retrieval characteristics of the {\tabjoinbench} subsets: number of unique queries, total number of relevant candidates, and average number of relevant candidates per query.}
\label{tab:retrieval_statistics}
\end{table}






Perturbations are applied to candidate tables only. This helps reduce lexical correspondence between the query and candidate tables. Candidate tables are perturbed to simulate the heterogeneity encountered in data lakes and to increase the number of relevant candidates for each query. The original candidate tables and perturbed variants are then added to the data lake. The resulting benchmark statistics are summarized in Tables~\ref{tab:benchmark_stats} and~\ref{tab:retrieval_statistics}. Segment-wise statistics for each perturbation category are provided in Appendix~\ref{appendix:perturbation_numbers}.

\begin{table*}[t]
\centering
\small
\setlength{\tabcolsep}{7.5pt}
\begin{tabular}{lccccccccc}
\toprule
Method & P@1 & R@1 & P@5 & R@5 & P@15 & R@15 & P@30 & R@30 & P@R \\
\midrule
\multicolumn{10}{c}{\textbf{(a) WikiTables Subset}} \\
\midrule
LSH Ensemble      & \textbf{0.969} & \textbf{0.038} & \textbf{0.959} & \textbf{0.178} & \textbf{0.950} & 0.462 & \textbf{0.937} & 0.516 & 0.529 \\
PEXESO            & 0.629 & 0.026 & 0.640 & 0.133 & 0.630 & 0.391 & 0.492 & 0.488 & 0.505 \\
D3L               & 0.199 & 0.007 & 0.175 & 0.026 & 0.150 & 0.063 & 0.125 & 0.089 & 0.101 \\
FREYJA            & 0.660 & 0.025 & 0.628 & 0.119 & 0.593 & 0.334 & 0.398 & 0.409 & 0.437 \\
WarpGate          & 0.220 & 0.007 & 0.205 & 0.034 & 0.177 & 0.085 & 0.140 & 0.123 & 0.133 \\
DeepJoin          & \underline{0.879} & \underline{0.034} & \underline{0.868} & \underline{0.167} & 0.836 & \textbf{0.485} & 0.661 & \textbf{0.726} & \textbf{0.762} \\
Qwen Emb. & 0.876 & \underline{0.034} & 0.836 & 0.159 & 0.795 & 0.451 & 0.629 & 0.678 & 0.688 \\
Gemma Emb. & 0.869 & 0.033 & 0.852 & 0.162 & \underline{0.840} & \underline{0.479} & \underline{0.668} & \underline{0.725} & \underline{0.749} \\
\midrule
\multicolumn{10}{c}{\textbf{(b) MMQA Subset}} \\
\midrule
LSH Ensemble & 0.161 & 0.006 & 0.160 & 0.030 & 0.159 & 0.086 & 0.156 & 0.142 & 0.137 \\
PEXESO 		 & 0.056 & 0.002 & 0.056 & 0.011 & 0.053 & 0.031 & 0.044 & 0.048 & 0.048 \\
D3L 		 & 0.309 & 0.012 & 0.288 & 0.057 & 0.265 & 0.156 & 0.232 & 0.265 & 0.241 \\
FREYJA 		 & 0.130 & 0.005 & 0.121 & 0.021 & 0.110 & 0.057 & 0.112 & 0.115 & 0.108 \\
WarpGate 	 & 0.282 & 0.011 & 0.286 & 0.055 & 0.233 & 0.130 & 0.208 & 0.229 & 0.208 \\
DeepJoin 	 & \textbf{0.585} & \textbf{0.023} & \textbf{0.549} & \textbf{0.107} & \textbf{0.519} & \textbf{0.307} & \textbf{0.446} & \textbf{0.513} & \textbf{0.478} \\
Qwen Emb. 	 & 0.446 & 0.017 & 0.403 & 0.078 & 0.360 & 0.209 & 0.296 & 0.336 & 0.304 \\
Gemma Emb. 	 & \underline{0.498} & \underline{0.019} & \underline{0.454} & \underline{0.088} & \underline{0.417} & \underline{0.241} & \underline{0.327} & \underline{0.369} & \underline{0.344} \\
\midrule
\multicolumn{10}{c}{\textbf{(c) NYC Open Data Subset}} \\
\midrule
LSH Ensemble & \underline{0.943} & \underline{0.025} & \textbf{0.937} & \textbf{0.126} & \textbf{0.938} & \textbf{0.379} & \textbf{0.764} & \underline{0.574} & 0.592 \\
PEXESO 		 & 0.317 & 0.009 & 0.313 & 0.042 & 0.314 & 0.127 & 0.239 & 0.193 & 0.199 \\
D3L 		 & 0.843 & 0.023 & 0.823 & 0.110 & 0.814 & 0.327 & 0.663 & 0.530 & 0.580 \\
FREYJA 		 & 0.929 & \underline{0.025} & \underline{0.923} & \underline{0.124} & \underline{0.894} & \underline{0.360} & \underline{0.751} & \textbf{0.595} & \textbf{0.655} \\
WarpGate 	 & 0.857 & 0.023 & 0.857 & 0.114 & 0.789 & 0.315 & 0.618 & 0.490 & 0.521 \\
DeepJoin 	 & \textbf{0.952} & \textbf{0.026} & 0.895 & 0.120 & 0.778 & 0.315 & 0.622 & 0.502 & 0.546 \\
Qwen Emb. 	 & 0.900 & 0.024 & 0.897 & 0.120 & 0.846 & 0.338 & 0.716 & 0.568 & \underline{0.644} \\
Gemma Emb. 	 & 0.800 & 0.021 & 0.797 & 0.105 & 0.769 & 0.303 & 0.678 & 0.534 & 0.617 \\
\bottomrule
\end{tabular}

\caption{Overall retrieval performance on the three {\tabjoinbench} subsets. The best and second-best results within each subset are shown in bold and underlined, respectively.}

\label{tab:overall-results}

\end{table*}

\section{Experiments}
\subsection{Experimental Setup}

\paragraph{Baselines.} We evaluate six representative methods spanning sketch-based retrieval, embedding-based similarity search, hybrid feature matching, profile-based prediction, learned table representations, alongside two general-purpose language-model embeddings. Specifically, we evaluate \textbf{LSH Ensemble} \cite{zhu_2016_lsh}, a sketch-based set-containment method; \textbf{PEXESO} \cite{dong_2021_pexeso}, an embedding-based similarity search method; \textbf{D3L} \cite{bogatu_2020_d3l}, a hybrid feature based approach; \textbf{Freyja} \cite{maynou_2026_freyja}, a profile-based join prediction method; \textbf{WarpGate} \cite{cong_2022_warpgate}, a learned table representation based approach; and \textbf{DeepJoin} \cite{dong_2023_deepjoin}, a learned language model based join discovery method. In addition, we evaluate language-model embedding baselines, \textbf{Qwen} (Qwen3-Embedding-0.6B) \cite{qwen3embedding} and \textbf{Gemma} (embeddinggemma-300m) \cite{embedding_gemma_2025}, using serialized column representations indexed with HNSW \cite{malkov_2020_hnsw}.

\paragraph{Evaluation Metrics.}
We report Precision@$k$ and Recall@$k$ for $k\in\{1,5,15,30\}$, along with Precision@$R$, where $R$ denotes the number of relevant candidates for a query. All metrics are computed per query and averaged across the subset.

\paragraph{Environment.}
All benchmark construction and evaluation code was implemented in Python~3.10. Experiments were conducted on an Ubuntu server equipped with two Intel Xeon processors (96 CPU cores), eight NVIDIA H200 GPUs, and 1.5~TB of memory; each execution used a single GPU. 

We use publicly available implementations of the evaluated methods where possible, including both author-released and third-party repositories. For methods without a suitable implementation, we implemented the methods following the original paper as closely as possible.

\paragraph{Evaluation Objectives.}
Our evaluation focuses on three aspects: (i) the performance of representative join discovery methods across the three benchmark settings, (ii) the effectiveness of language-model embeddings without join-specific training, and (iii) the robustness of these methods to structural, representation, and semantic perturbations.

\subsection{Results and Analysis}
Table~\ref{tab:overall-results} summarizes the retrieval performance of representative join discovery methods across the three \tabjoinbench{} benchmark subsets. 

\paragraph{WikiTables Subset.}
LSH Ensemble achieves the highest precision at each retrieval budget even though the method relies on sketch-based containment rather than learned semantic representations. DeepJoin obtains the highest P@R, indicating that it introduces the fewest irrelevant candidates within a query-specific retrieval budget. PEXESO and FREYJA achieve moderate performance, while D3L and WarpGate are substantially less effective on this subset.

\paragraph{MMQA Subset.}
DeepJoin consistently outperforms all other methods across every evaluation metric, achieving the highest precision, recall, and P@R. Among the specialized methods, D3L provides the strongest overall performance after DeepJoin, while PEXESO performs the weakest. Compared with the other benchmark subsets, all methods obtain substantially lower retrieval performance on MMQA Subset, indicating that it is the most challenging setting in \tabjoinbench. One possible explanation is that MMQA Subset primarily consists of relational joins over numeric identifier columns. Unlike textual attributes, these identifiers provide limited semantic context for embedding-based methods while simultaneously creating substantial overlap between candidate columns, resulting in a more challenging retrieval problem.

\paragraph{NYC Open Data Subset.}
Unlike the previous benchmark subsets, no single method consistently dominates on NYC Open Data. LSH Ensemble achieves the highest precision at retrieval budgets of 5, 15, and 30, while also ranking second on three additional metrics. DeepJoin obtains the highest P@1, indicating the strongest top-ranked retrieval, whereas FREYJA achieves the highest R@30 and P@R, supporting its design premise that profile-based representations can generalize across heterogeneous data lake collections. D3L also performs competitively on this subset, substantially outperforming its results on WikiTables and MMQA. In contrast, PEXESO remains the weakest method. One possible explanation is that the static GloVe embeddings used in our evaluation provide limited representations for the domain-specific entities commonly encountered in NYC Open Data.

\paragraph{Language-model Embedding Baselines.}
Across all three benchmark subsets, language-model embedding baselines perform competitively despite requiring no join-specific training. This indicates that advances in pretrained embedding models have substantially narrowed the gap between task-agnostic and specialized join discovery approaches.

\begin{table}[t]
\centering
\small
\setlength{\tabcolsep}{7.5pt}
\begin{tabular}{lccc}
\toprule
Method & Struct. & Rep. & Sem. \\
\midrule
\multicolumn{4}{c}{\textbf{WikiTables Subset}} \\
\midrule
LSH Ensemble & 0.749 & 0.482 & 0.534 \\
PEXESO       & 0.778 & 0.549 & 0.471 \\
D3L          & 0.150 & 0.109 & 0.090 \\
FREYJA       & 0.551 & 0.417 & 0.391 \\
WarpGate     & 0.156 & 0.150 & 0.103 \\
DeepJoin     & 0.807 & 0.808 & 0.786 \\
Qwen Emb.    & 0.749 & 0.747 & 0.696 \\
Gemma Emb.   & 0.815 & 0.808 & 0.788 \\
\midrule
\multicolumn{4}{c}{\textbf{MMQA Subset}} \\
\midrule
LSH Ensemble & 0.172 & 0.078 & 0.155 \\
PEXESO       & 0.064 & 0.041 & 0.059 \\
D3L          & 0.372 & 0.290 & 0.045 \\
FREYJA       & 0.140 & 0.084 & 0.105 \\
WarpGate     & 0.384 & 0.172 & 0.032 \\
DeepJoin     & 0.568 & 0.497 & 0.410 \\
Qwen Emb.    & 0.399 & 0.362 & 0.159 \\
Gemma Emb.   & 0.450 & 0.401 & 0.219 \\
\midrule
\multicolumn{4}{c}{\textbf{NYC Open Data Subset}} \\
\midrule
LSH Ensemble & 0.718 & 0.310 & 0.740 \\
PEXESO       & 0.313 & 0.123 & 0.249 \\
D3L          & 0.810 & 0.508 & 0.456 \\
FREYJA       & 0.703 & 0.520 & 0.753 \\
WarpGate     & 0.657 & 0.617 & 0.243 \\
DeepJoin     & 0.583 & 0.565 & 0.464 \\
Qwen Emb.    & 0.623 & 0.710 & 0.574 \\
Gemma Emb.   & 0.612 & 0.651 & 0.588 \\
\bottomrule
\end{tabular}
\caption{P@R under each perturbation category (Structural, Representation, Semantic) across \tabjoinbench{} subsets.}
\label{tab:perturbation_impact}
\end{table}

\paragraph{Impact of Perturbations.} Table~\ref{tab:perturbation_impact} reports P@R under each perturbation category (full metrics in in Appendix~\ref{appendix:perturbation_impact}). Perturbation sensitivity depends on the method as well as the underlying data source.

Value-overlap-, profile-, and sketch-based methods (PEXESO, FREYJA, LSH Ensemble) are degraded by representation perturbations on WikiTables and MMQA, since they depend on column values rather than names; this effect is strongest on MMQA, where numeric identifier keys carry little semantic content and lose value overlap directly under representation changes. It also explains why DeepJoin and language-model embedding baselines show a smaller advantage on MMQA. 

WikiTables shows a related but distinct pattern: because WikiTables tables are small and represent real-world entities, perturbations propagate across the entire table, and thus semantic perturbations are comparatively more disruptive than on other subsets, while structural perturbations have limited effect since row and column reorganization does not remove information from these tables.

NYC Open Data shows the opposite structural pattern. Because NYC candidates are constructed via vertical partitioning of a shared source table, structural perturbations directly alter candidate content and size, thus affecting value-overlap-, profile-, and sketch-based methods. While our semantic perturbations do modify values, these changes are applied non-exhaustively, and given NYC Open Data's larger value sets, sufficient lexical overlap typically survives for these methods to still detect a match. DeepJoin and embedding-based methods (Qwen, Gemma) reverse this pattern: these methods utilize column names as part of their representation and thus observe greater performance degradation under semantic perturbations, which target column names as well, than under structural changes, which leave names intact.

D3L and WarpGate are partial exceptions: although still affected by structural perturbations, both retain the Structural > Representation > Semantic ordering seen on WikiTables, MMQA, and NYC Open Data. D3L's reliance on column-name, format, and domain-distribution signals alongside values likely provides resilience to the value-overlap loss caused by structural perturbations, while WarpGate, as a learned table embedding method, likely retains some sensitivity to structural table organization learned during training, distinguishing it from the general-purpose language-model embeddings.


These results show that perturbation impact depends on both the table properties exploited by a join discovery method and the underlying data source, reinforcing that no single method or perturbation category dominates across settings.

\section{Related Work}
Research on join discovery has largely focused on developing retrieval methods, with comparatively less attention to benchmark construction. LakeBench \cite{srinivas_2023_lakebench, deng_2024_lakebench} is the closest work to {\tabjoinbench} and addresses scalable evaluation of join discovery methods. 
In contrast, {\tabjoinbench} focuses on constructing a benchmark spanning semantic, relational, and hybrid join settings.
MMTU \cite{xing2026mmtu} includes table join as one of several table understanding tasks, but does not evaluate join discovery as a retrieval task.
Several large-scale table corpora can serve as sources for benchmark construction, including GitTables \cite{hulsebos_2023_gittables}, SchemaPile \cite{dohmen_2024_schemapile}, and WikiDBs \cite{vogel2024wikidbs}.
However, these resources primarily support representation learning, schema understanding, and model pretraining, rather than retrieval-based benchmark evaluation.

\section{Conclusion}
We presented {\tabjoinbench}, a benchmark for evaluating join discovery methods across semantic, relational, and hybrid join settings. It supports a construction methodology to create a heterogeneous data lake through controlled structural, representation, and semantic perturbations while preserving reliable ground truth. Our evaluation of representative join discovery methods demonstrates that no single approach consistently dominates across all benchmark settings, highlighting the importance of diverse evaluation scenarios. We hope that {\tabjoinbench} provides a reproducible foundation for the development and evaluation of future join discovery methods.

\section{Limitations}
The construction of {\tabjoinbench} involves manual verification for the WikiTables and NYC Open Data subsets. This verification was performed by the primary author following a predefined annotation protocol. As a result, inter-annotator agreement could not be measured. To make manual verification tractable, source-specific ranking strategies were used to prioritize query--candidate pairs for inspection, and the final inclusion of pairs in the benchmark was determined by the annotator. Furthermore, {\tabjoinbench} provides verified positive join relationships but does not attempt to exhaustively enumerate all valid joins within the constructed data lake. Consequently, additional valid join relationships may exist between tables that are not included in the benchmark ground truth. Finally, the perturbation framework is designed to emulate realistic transformations observed during data integration, but it cannot capture their full diversity. Entity-aware perturbations are applied using predefined mappings and sampled LLM-assisted augmentations rather than exhaustively rewriting all applicable values. As a result, some query--candidate pairs may retain lexical or representation overlap despite the applied perturbations.

\section{Ethical Considerations}
{\tabjoinbench} is constructed from publicly available datasets, including WikiTables, MMQA, and NYC Open Data, and does not introduce any new personally identifiable information beyond that contained in the original sources. The released benchmark contains perturbed data generated solely for evaluating join discovery methods. These perturbed data should not be interpreted as authoritative or factual representations of the original datasets. Additionally, {\tabjoinbench} may reflect biases present in the source datasets. We used LLM assisted techniques to split files into semantically meaningful segments. For this purpose file metadata were exposed to LLM. Data were not exposed to LLM. AI writing assistants were used to polish the manuscript; all scientific claims, experimental design, and analysis were conducted and verified by the authors.

\bibliography{custom}

@article{zhu_2016_lsh,
author = {Zhu, Erkang and Nargesian, Fatemeh and Pu, Ken Q. and Miller, Ren\'{e}e J.},
title = {LSH ensemble: internet-scale domain search},
year = {2016},
issue_date = {August 2016},
publisher = {VLDB Endowment},
volume = {9},
number = {12},
issn = {2150-8097},
url = {https://doi.org/10.14778/2994509.2994534},
doi = {10.14778/2994509.2994534},
journal = {Proc. VLDB Endow.},
month = aug,
pages = {1185–1196},
numpages = {12}
}

@inproceedings{zhu_2019_josie,
author = {Zhu, Erkang and Deng, Dong and Nargesian, Fatemeh and Miller, Ren\'{e}e J.},
title = {JOSIE: Overlap Set Similarity Search for Finding Joinable Tables in Data Lakes},
year = {2019},
isbn = {9781450356435},
publisher = {Association for Computing Machinery},
address = {New York, NY, USA},
url = {https://doi.org/10.1145/3299869.3300065},
doi = {10.1145/3299869.3300065},
booktitle = {Proceedings of the 2019 International Conference on Management of Data},
pages = {847–864},
numpages = {18},
location = {Amsterdam, Netherlands},
series = {SIGMOD '19}
}

@INPROCEEDINGS{dong_2021_pexeso,
  author={Dong, Yuyang and Takeoka, Kunihiro and Xiao, Chuan and Oyamada, Masafumi},
  booktitle={2021 IEEE 37th International Conference on Data Engineering (ICDE)}, 
  title={Efficient Joinable Table Discovery in Data Lakes: A High-Dimensional Similarity-Based Approach}, 
  year={2021},
  volume={},
  number={},
  pages={456-467},
  doi={10.1109/ICDE51399.2021.00046}}

@INPROCEEDINGS{bogatu_2020_d3l,
  author={Bogatu, Alex and Fernandes, Alvaro A. A. and Paton, Norman W. and Konstantinou, Nikolaos},
  booktitle={2020 IEEE 36th International Conference on Data Engineering (ICDE)}, 
  title={Dataset Discovery in Data Lakes}, 
  year={2020},
  volume={},
  number={},
  pages={709-720},
  doi={10.1109/ICDE48307.2020.00067}}

@ARTICLE{maynou_2026_freyja,
  author={Maynou, Marc and Nadal, Sergi and Panadero, Raquel and Flores, Javier and Romero, Oscar and Queralt, Anna},
  journal={IEEE Transactions on Knowledge and Data Engineering}, 
  title={Freyja: Efficient Join Discovery in Data Lakes}, 
  year={2026},
  volume={38},
  number={4},
  pages={2277-2288},
  doi={10.1109/TKDE.2026.3656786}}

@article{cong_2022_warpgate,
  title={WarpGate: A Semantic Join Discovery System for Cloud Data Warehouses},
  author={Tianji Cong and James Gale and Jason Frantz and H. V. Jagadish and cCaugatay Demiralp},
  journal={ArXiv},
  year={2022},
  volume={abs/2212.14155},
  url={https://api.semanticscholar.org/CorpusID:255340476}
}

@article{dong_2023_deepjoin,
author = {Dong, Yuyang and Xiao, Chuan and Nozawa, Takuma and Enomoto, Masafumi and Oyamada, Masafumi},
title = {DeepJoin: Joinable Table Discovery with Pre-Trained Language Models},
year = {2023},
issue_date = {June 2023},
publisher = {VLDB Endowment},
volume = {16},
number = {10},
issn = {2150-8097},
url = {https://doi.org/10.14778/3603581.3603587},
doi = {10.14778/3603581.3603587},
journal = {Proc. VLDB Endow.},
month = jun,
pages = {2458–2470},
numpages = {13}
}

@article{koutras_2025_omnimatch,
author = {Koutras, Christos and Zhang, Jiani and Qin, Xiao and Lei, Chuan and Ioannidis, Vasileios and Faloutsos, Christos and Karypis, George and Katsifodimos, Asterios},
title = {OmniMatch: Joinability Discovery in Data Products},
year = {2025},
issue_date = {July 2025},
publisher = {VLDB Endowment},
volume = {18},
number = {11},
issn = {2150-8097},
url = {https://doi.org/10.14778/3749646.3749715},
doi = {10.14778/3749646.3749715},
journal = {Proc. VLDB Endow.},
month = jul,
pages = {4588–4601},
numpages = {14}
}

@INPROCEEDINGS{khatiwada_2025_tabsketchfm,
  author={Khatiwada, Aamod and Kokel, Harsha and Abdelaziz, Ibrahim and Chaudhury, Subhajit and Dolby, Julian and Hassanzadeh, Oktie and Huang, Zhenhan and Pedapati, Tejaswini and Samulowitz, Horst and Srinivas, Kavitha},
  booktitle={2025 IEEE 41st International Conference on Data Engineering (ICDE)}, 
  title={TabSketchFM: Sketch-Based Tabular Representation Learning for Data Discovery Over Data Lakes}, 
  year={2025},
  volume={},
  number={},
  pages={1523-1536},
  doi={10.1109/ICDE65448.2025.00118}}

@inproceedings{lehmberg_2016_wettables,
author = {Lehmberg, Oliver and Ritze, Dominique and Meusel, Robert and Bizer, Christian},
title = {A Large Public Corpus of Web Tables containing Time and Context Metadata},
year = {2016},
isbn = {9781450341448},
publisher = {International World Wide Web Conferences Steering Committee},
address = {Republic and Canton of Geneva, CHE},
url = {https://doi.org/10.1145/2872518.2889386},
doi = {10.1145/2872518.2889386},
booktitle = {Proceedings of the 25th International Conference Companion on World Wide Web},
pages = {75–76},
numpages = {2},
location = {Montr\'{e}al, Qu\'{e}bec, Canada},
series = {WWW '16 Companion}
}

@InProceedings{bhagavatula_2015_wikitables,
author="Bhagavatula, Chandra Sekhar
and Noraset, Thanapon
and Downey, Doug",
editor="Arenas, Marcelo
and Corcho, Oscar
and Simperl, Elena
and Strohmaier, Markus
and d'Aquin, Mathieu
and Srinivas, Kavitha
and Groth, Paul
and Dumontier, Michel
and Heflin, Jeff
and Thirunarayan, Krishnaprasad
and Thirunarayan, Krishnaprasad
and Staab, Steffen",
title="TabEL: Entity Linking in Web Tables",
booktitle="The Semantic Web - ISWC 2015",
year="2015",
publisher="Springer International Publishing",
address="Cham",
pages="425--441",
isbn="978-3-319-25007-6",
url={http://websail-fe.cs.northwestern.edu/TabEL/}
}

@misc{srinivas_2023_lakebench,
      title={LakeBench: Benchmarks for Data Discovery over Data Lakes}, 
      author={Kavitha Srinivas and Julian Dolby and Ibrahim Abdelaziz and Oktie Hassanzadeh and Harsha Kokel and Aamod Khatiwada and Tejaswini Pedapati and Subhajit Chaudhury and Horst Samulowitz},
      year={2023},
      eprint={2307.04217},
      archivePrefix={arXiv},
      primaryClass={cs.DB},
      url={https://arxiv.org/abs/2307.04217}, 
}

@article{deng_2024_lakebench,
author = {Deng, Yuhao and Chai, Chengliang and Cao, Lei and Yuan, Qin and Chen, Siyuan and Yu, Yanrui and Sun, Zhaoze and Wang, Junyi and Li, Jiajun and Cao, Ziqi and Jin, Kaisen and Zhang, Chi and Jiang, Yuqing and Zhang, Yuanfang and Wang, Yuping and Yuan, Ye and Wang, Guoren and Tang, Nan},
title = {LakeBench: A Benchmark for Discovering Joinable and Unionable Tables in Data Lakes},
year = {2024},
issue_date = {April 2024},
publisher = {VLDB Endowment},
volume = {17},
number = {8},
issn = {2150-8097},
url = {https://doi.org/10.14778/3659437.3659448},
doi = {10.14778/3659437.3659448},
journal = {Proc. VLDB Endow.},
month = apr,
pages = {1925–1938},
numpages = {14}
}

@inproceedings{
wu2025mmqa,
title={{MMQA}: Evaluating {LLM}s with Multi-Table Multi-Hop Complex Questions},
author={Jian Wu and Linyi Yang and Dongyuan Li and Yuliang Ji and Manabu Okumura and Yue Zhang},
booktitle={The Thirteenth International Conference on Learning Representations},
year={2025},
url={https://openreview.net/forum?id=GGlpykXDCa}
}

@inproceedings{yu-etal-2018-spider,
    title = "{S}pider: A Large-Scale Human-Labeled Dataset for Complex and Cross-Domain Semantic Parsing and Text-to-{SQL} Task",
    author = "Yu, Tao  and
      Zhang, Rui  and
      Yang, Kai  and
      Yasunaga, Michihiro  and
      Wang, Dongxu  and
      Li, Zifan  and
      Ma, James  and
      Li, Irene  and
      Yao, Qingning  and
      Roman, Shanelle  and
      Zhang, Zilin  and
      Radev, Dragomir",
    editor = "Riloff, Ellen  and
      Chiang, David  and
      Hockenmaier, Julia  and
      Tsujii, Jun{'}ichi",
    booktitle = "Proceedings of the 2018 Conference on Empirical Methods in Natural Language Processing",
    month = oct # "-" # nov,
    year = "2018",
    address = "Brussels, Belgium",
    publisher = "Association for Computational Linguistics",
    url = "https://aclanthology.org/D18-1425/",
    doi = "10.18653/v1/D18-1425",
    pages = "3911--3921"
}

@article{hulsebos_2023_gittables,
author = {Hulsebos, Madelon and Demiralp, \c{C}agatay and Groth, Paul},
title = {GitTables: A Large-Scale Corpus of Relational Tables},
year = {2023},
issue_date = {May 2023},
publisher = {Association for Computing Machinery},
address = {New York, NY, USA},
volume = {1},
number = {1},
url = {https://doi.org/10.1145/3588710},
doi = {10.1145/3588710},
journal = {Proc. ACM Manag. Data},
month = may,
articleno = {30},
numpages = {17}
}

@article{dohmen_2024_schemapile,
author = {D\"{o}hmen, Till and Geacu, Radu and Hulsebos, Madelon and Schelter, Sebastian},
title = {SchemaPile: A Large Collection of Relational Database Schemas},
year = {2024},
issue_date = {June 2024},
publisher = {Association for Computing Machinery},
address = {New York, NY, USA},
volume = {2},
number = {3},
url = {https://doi.org/10.1145/3654975},
doi = {10.1145/3654975},
journal = {Proc. ACM Manag. Data},
month = may,
articleno = {172},
numpages = {25}
}

@inproceedings{
vogel2024wikidbs,
title={Wiki{DB}s: A Large-Scale Corpus Of Relational Databases From Wikidata},
author={Liane Vogel and Jan-Micha Bodensohn and Carsten Binnig},
booktitle={The Thirty-eight Conference on Neural Information Processing Systems Datasets and Benchmarks Track},
year={2024},
url={https://openreview.net/forum?id=abXaOcvujs}
}

@inproceedings{
xing2026mmtu,
title={{MMTU}: A Massive Multi-Task Table Understanding and Reasoning Benchmark},
author={Junjie Xing and Yeye He and Mengyu Zhou and Haoyu Dong and Shi Han and Lingjiao Chen and Dongmei Zhang and Surajit Chaudhuri and H. V. Jagadish},
booktitle={The Thirty-ninth Annual Conference on Neural Information Processing Systems Datasets and Benchmarks Track},
year={2026},
url={https://openreview.net/forum?id=ryUzgwD6UQ}
}

@article{navathe_1984_vp,
author = {Navathe, Shamkant and Ceri, Stefano and Wiederhold, Gio and Dou, Jinglie},
title = {Vertical partitioning algorithms for database design},
year = {1984},
issue_date = {Dec. 1984},
publisher = {Association for Computing Machinery},
address = {New York, NY, USA},
volume = {9},
number = {4},
issn = {0362-5915},
url = {https://doi.org/10.1145/1994.2209},
doi = {10.1145/1994.2209},
journal = {ACM Trans. Database Syst.},
month = dec,
pages = {680–710},
numpages = {31}
}

@article{qwen3embedding,
  title={Qwen3 Embedding: Advancing Text Embedding and Reranking Through Foundation Models},
  author={Zhang, Yanzhao and Li, Mingxin and Long, Dingkun and Zhang, Xin and Lin, Huan and Yang, Baosong and Xie, Pengjun and Yang, An and Liu, Dayiheng and Lin, Junyang and Huang, Fei and Zhou, Jingren},
  journal={arXiv preprint arXiv:2506.05176},
  year={2025}
}

@article{embedding_gemma_2025,
    title={EmbeddingGemma: Powerful and Lightweight Text Representations},
    author={Schechter Vera, Henrique* and Dua, Sahil* and Zhang, Biao and Salz, Daniel and Mullins, Ryan and Raghuram Panyam, Sindhu and Smoot, Sara and Naim, Iftekhar and Zou, Joe and Chen, Feiyang and Cer, Daniel and Lisak, Alice and Choi, Min and Gonzalez, Lucas and Sanseviero, Omar and Cameron, Glenn and Ballantyne, Ian and Black, Kat and Chen, Kaifeng and Wang, Weiyi and Li, Zhe and Martins, Gus and Lee, Jinhyuk and Sherwood, Mark and Ji, Juyeong and Wu, Renjie and Zheng, Jingxiao and Singh, Jyotinder and Sharma, Abheesht and Sreepat, Divya and Jain, Aashi and Elarabawy, Adham and Co, AJ and Doumanoglou, Andreas and Samari, Babak and Hora, Ben and Potetz, Brian and Kim, Dahun and Alfonseca, Enrique and Moiseev, Fedor and Han, Feng and Palma Gomez, Frank and Hernández Ábrego, Gustavo and Zhang, Hesen and Hui, Hui and Han, Jay and Gill, Karan and Chen, Ke and Chen, Koert and Shanbhogue, Madhuri and Boratko, Michael and Suganthan, Paul and Duddu, Sai Meher Karthik and Mariserla, Sandeep and Ariafar, Setareh and Zhang, Shanfeng and Zhang, Shijie and Baumgartner, Simon and Goenka, Sonam and Qiu, Steve and Dabral, Tanmaya and Walker, Trevor and Rao, Vikram and Khawaja, Waleed and Zhou, Wenlei and Ren, Xiaoqi and Xia, Ye and Chen, Yichang and Chen, Yi-Ting and Dong, Zhe and Ding, Zhongli and Visin, Francesco and Liu, Gaël and Zhang, Jiageng and Kenealy, Kathleen and Casbon, Michelle and Kumar, Ravin and Mesnard, Thomas and Gleicher, Zach and Brick, Cormac and Lacombe, Olivier and Roberts, Adam and Sung, Yunhsuan and Hoffmann, Raphael and Warkentin, Tris and Joulin, Armand and Duerig, Tom and Seyedhosseini, Mojtaba},
    publisher={Google DeepMind},
    year={2025},
    url={https://arxiv.org/abs/2509.20354},
    journal={arXiv preprint arXiv:2509.20354}
}

@article{malkov_2020_hnsw,
author = {Malkov, Yu A. and Yashunin, D. A.},
title = {Efficient and Robust Approximate Nearest Neighbor Search Using Hierarchical Navigable Small World Graphs},
year = {2020},
issue_date = {April 2020},
publisher = {IEEE Computer Society},
address = {USA},
volume = {42},
number = {4},
issn = {0162-8828},
url = {https://doi.org/10.1109/TPAMI.2018.2889473},
doi = {10.1109/TPAMI.2018.2889473},
journal = {IEEE Trans. Pattern Anal. Mach. Intell.},
month = apr,
pages = {824–836},
numpages = {13}
}

\appendix
\section{Annotation Protocol}
\label{appendix:protocol}
\subsection{WikiTables Annotation}

Each candidate query--table pair was assigned one of four labels: \textit{Joinable}, \textit{Unionable}, \textit{Duplicate/Subset}, or \textit{Undetermined}. To prioritize efficient triage toward high-quality positive pairs, pairs without a clear, immediately identifiable relationship were set aside under the \textit{Undetermined} label rather than analyzed further.

The following guidelines were used:
\begin{itemize}
    \item Prefer joins that provide complementary information rather than duplicate existing attributes.
    \item Accept joins based on meaningful entities (e.g., person, organization, team, location, institution, chemical compound, political party).
    \item Treat joins relying only on generic attributes (e.g., year, season, event, category, or common column names) as \textit{Undetermined}, unless they uniquely identify the same real-world entity.
    \item Treat pairs representing different events, time periods, geographic regions, or incompatible levels of granularity as \textit{Undetermined}.
    \item If a candidate is primarily unionable or duplicate rather than complementary, label it \textit{Unionable} or \textit{Duplicate/Subset}, not \textit{Joinable}.
\end{itemize}

Only \textit{Joinable} pairs were retained as seed pairs.

\subsection{NYC Open Data Validation}

For NYC Open Data, the query and candidate tables originate from the same source dataset. Manual validation therefore focused on verifying the quality of the generated partition rather than the existence of a join. The following guidelines were applied:

\begin{itemize}
    \item Verify that the generated column groups represent coherent real-world entities.
    \item Verify that the shared column is a meaningful join attribute.
    \item Prefer descriptive textual attributes over identifier or code columns when both represent the same entity.
    \item Avoid selecting excessively short categorical attributes or long free-text descriptive fields as join columns.
    \item Treat partitions that do not produce meaningful complementary tables as \textit{Undetermined}.
\end{itemize}

Only validated partitions were retained as benchmark instances.

\section{NYC Open Data Partitioning Prompt}
\label{appendix:prompt}
The following prompt was used with GPT-4o to generate the initial partitioning of NYC Open Data datasets into two normalized column groups.

\begin{lstlisting}[basicstyle=\ttfamily\small,frame=single,breaklines=true]
You are an expert database designer.

Given:

1. A dataset description.
2. The column names.

Your task is to perform the first step of database normalization.

Instructions:

1. Identify the primary real-world entities represented by the dataset.
2. Infer the most likely natural keys (Use a column strictly from the column list. Do not invent surrogate keys). When the entity has both a coded/abbreviated identifier and a descriptive, human-readable column name that identifies the same real-world entity, prefer the descriptive name as the shared column instead of the code. A descriptive column is human-readable and refers to a real-world entity by name; a coded column is a short, abbreviated identifier with no inherent meaning without a lookup. Fall back to the coded identifier if no descriptive alternative exists in the column list.
3. Identify likely functional dependencies between columns based on their names, and dataset description.
4. Partition the columns into exactly two normalized groups (tables).
5. Exactly ONE column should appear in both groups because it represents the relationship between the two entities.
6. Every other column must belong to exactly one group.
7. Minimize redundancy.
8. Maximize functional dependency within each group.
9. The groups should correspond to meaningful entities.
10. Give a short but meaningful name for each partition representing the entity.
11. For every column, write a short (max one line) description of what that column represents, specific to this dataset.

Return the answer as JSON.

Output format:

{
    "group_1": "GroupName",
    "column_group_1": ["shared_column", "column11", "column12", ... ],
    "group_2": "GroupName",
    "column_group_2": ["shared_column", "column21", "column22", ... ],
    "shared_column": "shared_column",
    "column_description": {"column1": "short description", "column2": "short description", ... }
}
\end{lstlisting}

\section{Composition of Perturbed Segments}
\label{appendix:perturbation_numbers}
Table~\ref{tab:perturbation_type_stat} reports the number of relevant candidate pairs and average relevant candidates per query for each perturbation category and subset.

\begin{table}[!ht]
\centering
\small
\begin{tabular}{llcc}
\toprule
\makecell[l]{\textbf{Subset}} &
\makecell[l]{\textbf{Perturbation}\\\textbf{Type}} &
\makecell[c]{\textbf{Relevant}\\\textbf{Pairs}} &
\makecell[c]{\textbf{Relevant}\\\textbf{/ Query}} \\
\midrule
\multirow{4}{*}{\shortstack[l]{WikiTables\\(295 Queries)}}
 & Structural     & 3,995  & 13.54 \\
 & Representation & 3,677  & 12.46 \\
 & Semantic       & 4,244  & 14.39 \\
 \cmidrule(lr){2-4}
 & \textit{Total} & \textit{11,916} & \textit{40.39} \\
\midrule
\multirow{4}{*}{\shortstack[l]{MMQA\\(408 Queries)}}
 & Structural     & 4,586  & 11.24 \\
 & Representation & 4,195  & 10.28 \\
 & Semantic       & 4,409  & 10.81 \\
 \cmidrule(lr){2-4}
 & \textit{Total} & \textit{13,190} & \textit{32.33} \\
\midrule
\multirow{4}{*}{\shortstack[l]{NYC \\Open Data\\(70 Queries)}}
 & Structural     & 884    & 12.63 \\
 & Representation & 854    & 12.20 \\
 & Semantic       & 900    & 12.86 \\
 \cmidrule(lr){2-4}
 & \textit{Total} & \textit{2,638}  & \textit{37.69} \\
\bottomrule
\end{tabular}
\caption{Number of relevant candidate pairs and average relevant candidates per query, by perturbation category, across \tabjoinbench{} subsets.}
\label{tab:perturbation_type_stat}
\end{table}

\section{Extended Perturbation Results}
\label{appendix:perturbation_impact}
Table~\ref{tab:wiki_perturb}, Table~\ref{tab:mmqa_perturb}, and Table~\ref{tab:nyc_perturb} show the impact of perturbations on the WikiTables, MMQA, and NYC Open Data subsets, respectively. We report $k \in \{1, 5, 15\}$, since each perturbation segment contains at most 14 relevant candidates per query on average.

\begin{table}[!ht]
\centering
\scriptsize
\setlength{\tabcolsep}{2.5pt}
\begin{tabular}{lcccccc}
\toprule
Method & P@1 & R@1 & P@5 & R@5 & P@15 & R@15\\
\midrule
\multicolumn{7}{c}{\textbf{Structural Perturbations}} \\
\midrule
LSH Ensemble & 0.947 & 0.125 & 0.956 & 0.621 & 0.936 & 0.744 \\
PEXESO & 0.784 & 0.107 & 0.793 & 0.541 & 0.571 & 0.767 \\
D3L & 0.168 & 0.017 & 0.166 & 0.085 & 0.124 & 0.152 \\
FREYJA & 0.625 & 0.080 & 0.609 & 0.389 & 0.340 & 0.545 \\
WarpGate & 0.190 & 0.019 & 0.172 & 0.088 & 0.129 & 0.173 \\
DeepJoin & 0.857 & 0.111 & 0.837 & 0.546 & 0.496 & 0.812 \\
Qwen Emb. & 0.821 & 0.107 & 0.791 & 0.509 & 0.485 & 0.787 \\
Gemma Emb. & 0.856 & 0.111 & 0.850 & 0.550 & 0.510 & 0.833 \\
\midrule
\multicolumn{7}{c}{\textbf{Representation Perturbations}} \\
\midrule
LSH Ensemble & 0.945 & 0.115 & 0.938 & 0.434 & 0.944 & 0.495 \\
PEXESO & 0.730 & 0.090 & 0.724 & 0.445 & 0.493 & 0.579 \\
D3L & 0.210 & 0.022 & 0.170 & 0.080 & 0.117 & 0.113 \\
FREYJA & 0.643 & 0.076 & 0.592 & 0.349 & 0.277 & 0.419 \\
WarpGate & 0.221 & 0.024 & 0.195 & 0.104 & 0.124 & 0.168 \\
DeepJoin & 0.879 & 0.107 & 0.857 & 0.518 & 0.516 & 0.828 \\
Qwen Emb. & 0.864 & 0.104 & 0.817 & 0.490 & 0.485 & 0.785 \\
Gemma Emb. & 0.867 & 0.105 & 0.858 & 0.520 & 0.510 & 0.837 \\
\midrule
\multicolumn{7}{c}{\textbf{Semantic Perturbations}} \\
\midrule
LSH Ensemble & 0.973 & 0.108 & 0.961 & 0.461 & 0.940 & 0.525 \\
PEXESO & 0.693 & 0.078 & 0.661 & 0.362 & 0.430 & 0.470 \\
D3L & 0.178 & 0.018 & 0.122 & 0.051 & 0.097 & 0.091 \\
FREYJA & 0.654 & 0.070 & 0.568 & 0.298 & 0.291 & 0.386 \\
WarpGate & 0.231 & 0.022 & 0.161 & 0.070 & 0.100 & 0.110 \\
DeepJoin & 0.901 & 0.103 & 0.848 & 0.458 & 0.555 & 0.787 \\
Qwen Emb. & 0.860 & 0.094 & 0.792 & 0.414 & 0.520 & 0.730 \\
Gemma Emb. & 0.892 & 0.097 & 0.853 & 0.449 & 0.564 & 0.790 \\
\bottomrule
\end{tabular}
\caption{Retrieval performance on the perturbed segments of WikiTables Subset.}
\label{tab:wiki_perturb}
\end{table}

\begin{table}[ht]
\centering
\scriptsize
\setlength{\tabcolsep}{2.5pt}
\begin{tabular}{lcccccc}
\toprule
Method & P@1 & R@1 & P@5 & R@5 & P@15 & R@15\\
\midrule
\multicolumn{7}{c}{\textbf{Structural Perturbations}} \\
\midrule
LSH Ensemble & 0.177 & 0.022 & 0.174 & 0.105 & 0.172 & 0.225 \\
PEXESO & 0.067 & 0.013 & 0.062 & 0.055 & 0.050 & 0.098 \\
D3L & 0.341 & 0.044 & 0.362 & 0.231 & 0.289 & 0.503 \\
FREYJA & 0.142 & 0.017 & 0.139 & 0.084 & 0.125 & 0.212 \\
WarpGate & 0.324 & 0.041 & 0.376 & 0.239 & 0.300 & 0.523 \\
DeepJoin & 0.592 & 0.071 & 0.563 & 0.340 & 0.449 & 0.749 \\
Qwen Emb. & 0.444 & 0.056 & 0.413 & 0.260 & 0.314 & 0.547 \\
Gemma Emb. & 0.495 & 0.061 & 0.465 & 0.289 & 0.334 & 0.571 \\
\midrule
\multicolumn{7}{c}{\textbf{Representation Perturbations}} \\
\midrule
LSH Ensemble & 0.102 & 0.011 & 0.095 & 0.044 & 0.115 & 0.112 \\
PEXESO & 0.044 & 0.004 & 0.044 & 0.022 & 0.044 & 0.050 \\
D3L & 0.309 & 0.039 & 0.286 & 0.175 & 0.260 & 0.455 \\
FREYJA & 0.118 & 0.012 & 0.091 & 0.046 & 0.084 & 0.121 \\
WarpGate & 0.184 & 0.022 & 0.179 & 0.108 & 0.177 & 0.304 \\
DeepJoin & 0.531 & 0.066 & 0.502 & 0.312 & 0.396 & 0.702 \\
Qwen Emb. & 0.387 & 0.047 & 0.367 & 0.221 & 0.297 & 0.516 \\
Gemma Emb. & 0.439 & 0.053 & 0.414 & 0.247 & 0.312 & 0.534 \\
\midrule
\multicolumn{7}{c}{\textbf{Semantic Perturbations}} \\
\midrule
LSH Ensemble & 0.164 & 0.019 & 0.170 & 0.095 & 0.167 & 0.205 \\
PEXESO & 0.078 & 0.007 & 0.067 & 0.034 & 0.053 & 0.065 \\
D3L & 0.076 & 0.009 & 0.045 & 0.025 & 0.038 & 0.06 \\
FREYJA & 0.113 & 0.012 & 0.103 & 0.053 & 0.108 & 0.159 \\
WarpGate & 0.076 & 0.008 & 0.030 & 0.015 & 0.028 & 0.042 \\
DeepJoin & 0.500 & 0.058 & 0.431 & 0.252 & 0.345 & 0.579 \\
Qwen Emb. & 0.248 & 0.028 & 0.173 & 0.097 & 0.146 & 0.240 \\
Gemma Emb. & 0.321 & 0.037 & 0.248 & 0.141 & 0.192 & 0.319 \\
\bottomrule
\end{tabular}
\caption{Retrieval performance on the perturbed segments of MMQA Subset.}
\label{tab:mmqa_perturb}
\end{table}

\begin{table}[ht]
\centering
\scriptsize
\setlength{\tabcolsep}{2.5pt}
\begin{tabular}{lcccccc}
\toprule
Method & P@1 & R@1 & P@5 & R@5 & P@15 & R@15\\
\midrule
\multicolumn{7}{c}{\textbf{Structural Perturbations}} \\
\midrule
LSH Ensemble & 0.943 & 0.075 & 0.943 & 0.375 & 0.728 & 0.725 \\
PEXESO & 0.367 & 0.029 & 0.36 & 0.141 & 0.289 & 0.325 \\
D3L & 0.843 & 0.067 & 0.851 & 0.339 & 0.719 & 0.827 \\
FREYJA & 0.900 & 0.071 & 0.897 & 0.353 & 0.620 & 0.729 \\
WarpGate & 0.871 & 0.069 & 0.700 & 0.276 & 0.586 & 0.690 \\
DeepJoin & 0.905 & 0.071 & 0.667 & 0.260 & 0.540 & 0.628 \\
Qwen Emb. & 0.900 & 0.071 & 0.694 & 0.273 & 0.565 & 0.666 \\
Gemma Emb. & 0.800 & 0.064 & 0.686 & 0.270 & 0.552 & 0.650 \\
\midrule
\multicolumn{7}{c}{\textbf{Representation Perturbations}} \\
\midrule
LSH Ensemble & 0.643 & 0.052 & 0.589 & 0.228 & 0.530 & 0.340 \\
PEXESO & 0.200 & 0.016 & 0.177 & 0.070 & 0.113 & 0.131 \\
D3L & 0.786 & 0.066 & 0.729 & 0.303 & 0.472 & 0.526 \\
FREYJA & 0.886 & 0.075 & 0.763 & 0.313 & 0.439 & 0.530 \\
WarpGate & 0.886 & 0.075 & 0.831 & 0.343 & 0.519 & 0.635 \\
DeepJoin & 0.905 & 0.173 & 0.771 & 0.325 & 0.470 & 0.597 \\
Qwen Emb. & 0.929 & 0.078 & 0.889 & 0.368 & 0.602 & 0.736 \\
Gemma Emb. & 0.800 & 0.066 & 0.783 & 0.322 & 0.574 & 0.700 \\
\midrule
\multicolumn{7}{c}{\textbf{Semantic Perturbations}} \\
\midrule
LSH Ensemble & 0.929 & 0.076 & 0.937 & 0.380 & 0.761 & 0.739 \\
PEXESO & 0.317 & 0.027 & 0.307 & 0.128 & 0.213 & 0.255 \\
D3L & 0.671 & 0.053 & 0.609 & 0.238 & 0.433 & 0.470 \\
FREYJA & 0.900 & 0.073 & 0.909 & 0.366 & 0.645 & 0.756 \\
WarpGate & 0.671 & 0.051 & 0.500 & 0.182 & 0.228 & 0.247 \\
DeepJoin & 0.762 & 0.059 & 0.629 & 0.242 & 0.432 & 0.498 \\
Qwen Emb. & 0.771 & 0.06 & 0.686 & 0.26 & 0.525 & 0.581 \\
Gemma Emb. & 0.743 & 0.057 & 0.674 & 0.254 & 0.534 & 0.595 \\
\bottomrule
\end{tabular}
\caption{Retrieval performance on the perturbed segments of NYC Open Data Subset.}
\label{tab:nyc_perturb}
\end{table}

\section{Method Parameters}
\label{appendix:method_parameters}

Unless otherwise specified, all methods were evaluated using the default hyperparameters provided by their original implementations. Method-specific parameters are listed below.

\paragraph{PEXESO.}
PEXESO used the \texttt{glove-wiki-gigaword-50} embedding model. For the WikiTables subset, the indexing parameters were \textit{pivot}=3, \textit{base}=2, \textit{level}=4, $\tau=0.6$, and $T=0.4$. For the MMQA and NYC Open Data subsets, the parameters were \textit{pivot}=5, \textit{base}=2, \textit{level}=6, $\tau=0.6$, and $T=0.4$.

\paragraph{LSH Ensemble.}
The index was constructed using 256 hash functions and 32 partitions, with a containment threshold of 0.8.

\paragraph{D3L.}
The similarity threshold was set to 0.5. Hash sizes of 256 were used for the name, format, and value indices, while a hash size of 1024 was used for the embedding and distribution indices. The q-gram size was set to 3.

\paragraph{WarpGate.}
WarpGate used the \texttt{tabert\_large\_k3} table embedder.

\paragraph{DeepJoin.}
DeepJoin was trained using \texttt{MultipleNegativesRankingLoss}. Training used a batch size of 32 for 4 epochs with a 10\% validation split, 10,000 warm-up steps, and a weight decay of 0.01.

\paragraph{Language Model Embeddings.}
Column embeddings were indexed using an HNSW index with cosine similarity, $ef_{\mathrm{construction}}=100$, and $M=32$. The embedding dimension was determined by the selected embedding model.

\end{document}